\documentstyle[prl,preprint,epsfig,aps]{revtex}

\newcommand{\nn}{\nonumber}
\newcommand{\eea}{\end{eqnarray}}
\newcommand{\bea}{\begin{eqnarray}}
\newcommand{\ba}{\begin{array}}
\newcommand{\ea}{\end{array}}
\newcommand{\be}{\begin{equation}}
\newcommand{\ee}{\end{equation}}
\newcommand{\bd}{\begin{description}} \newcommand{\ed}{\end{description}}
\newcommand{\bfig}{\begin{figure}} \newcommand{\efig}{\end{figure}}

\begin{document}
\draft \preprint{HEP/123-qed}
 \tightenlines
\title{Modified Renormalization Strategy for Sandpile Models}
\author{Y. Moreno$^1$\cite{byline}, J. B. G\'{o}mez$^2$, A. F. Pacheco$^1$}
\address{
$^1$ Departamento de F\'{\i}sica Te\'{o}rica, Universidad de
Zaragoza, 50009 Zaragoza, Spain.\\ $^2$ Departamento de Ciencias
de la Tierra,
 Universidad de Zaragoza, 50009 Zaragoza, Spain.}
 \date{\today}
 \maketitle
 \widetext
\begin{abstract} Following the Renormalization Group scheme recently developed by
Pietronero {\it et al}, we introduce a simplifying strategy for the renormalization of
the relaxation dynamics of sandpile models. In our scheme, five sub-cells at a generic
scale $b$ form the renormalized cell at the next larger scale. Now the fixed point has a
unique nonzero dynamical component that allows for a great simplification in the
computation of the critical exponent $z$. The values obtained are in good agreement with
both numerical and theoretical results previously reported. \end{abstract} 

%\begin{multicols}{2}
%\narrowtext

%\section{Introduction}
%\label{sec:level1}
The concept of Self-Organized Criticality (SOC) introduced by Bak,
Tang and Wiesenfeld (BTW) \cite{bak88} has attracted a wide
interest to understand a class of dynamically driven systems which
self-organize into a statistically stationary state characterized
by the lack of any typical time or length scale. Numerical results
for systems displaying SOC behavior have been extensively reported
\cite{nsm,jensen}, but only a few theoretical approaches are known
to be in agreement with numerical simulations in all dimensions.
The major source of difficulties in the study of SOC systems lies
in their inherent complexity that makes the models analytically
tractable only in a few cases. The Abelian version of the BTW
sandpile model, early addressed by Dhar \cite{dhar}, turned out to
be one of these exceptions.

Recently, Pietronero, Vespignani, and Zapperi \cite{pietro94}
developed a new type of real space renormalization group approach
for dynamically driven systems, able to describe the
self-organized critical state of sandpile models by defining a
characterization of the phase space in which it is possible the
renormalization of the dynamics under repeated change of scale. In
addition, it is also possible to compute the critical exponents
analytically \cite{vesp95}. The method also reveals the nature of
the SOC problems and provides a picture about the universality
classes of different sandpile models. This scheme of
renormalization has been recently improved by considering
increasingly complex proliferation paths \cite{iv,ipvz} and
extended to forest-fire models \cite{lvz,vzl,lpvz}.

In this report, we follow the renormalization procedure of references
\cite{pietro94,vesp95} but using a Greek cross-shaped cell instead of a square cell in
the renormalization of the relaxation dynamics. The critical exponents that characterize
the stationary state are then computed and they are found to be in good agreement with
previous theoretical results and large scale numerical simulations both for the BTW and
two state model of Manna. We will see that the use of this particular choice of cells
simplifies the renormalization equations for the BTW model.

In what follows, we will focus on the sandpile critical height models in two dimensions.
Sandpile models are cellular automaton defined on a lattice where to each site one
assigns a variable (to which we will refer as energy). We let the system evolves by
randomly adding units of energy on the system. When the energy of a site reaches a
critical value, it relaxes releasing its entire energy to the neighboring sites. The
affected sites may become unstable triggering new toppling events and so on until all
sites are again stable. Three different classes of sites can be distinguished: (i) those
sites for which the addition of a unit of energy does not induce relaxation (stable
sites), (ii) those sites for which the addition of a unit of energy provokes they become
unstable (critical sites), and (iii) unstable sites that will relax at the next time
step. Open boundary conditions allow the energy to leave the system.

In this formalism, we will denote by $\rho$ the density of
critical sites. These definitions can be extended to a generic
scale $b$ by considering coarse grained variables. Thus, a cell at
scale $b$ is considered critical if the addition of a unit of
energy $\delta E(b)$ induces a relaxation of the size of the cell,
that is, the subrelaxation processes span the cell and transfer
energy to some neighbors. According to \cite{pietro94} the
relaxation process can lead to four different possibilities at
coarse grained levels: the energy can be distributed to one, two,
three, or four neighbors with probabilities $p_1$, $p_2$, $p_3$,
and $p_4$ respectively. Of course, it is also possible that in
certain cases the unstable sites at coarse grained scale do not
transfer energy to their nearest neighbors as well as to consider
different proliferation problems. We, as in
\cite{pietro94,vesp95}, will not consider these cases
\cite{note1}. Then, the probability distribution is defined by the
vector
 \be
 \overrightarrow{P}=(p_1,p_2,p_3,p_4)
 \ee
 with the normalization condition $\sum^{4}_{i=1}p_i=1$.

So, the properties of the system are fully characterized by the
distribution $(\rho,\overrightarrow{P})$ at each scale. The
relation between $\rho$ and $\overrightarrow{P}$ can be derived by
noting that in the stationary state the inflow of energy equals
the flow of energy out of the system \cite{com1}. This implies
\cite{vesp95}:
 \be
 \rho^{(k)}=\frac{1}{\sum_{i} ip_i^{(k)}}\quad , \label{two}
 \ee
which allow us to evaluate the stationary distribution of critical
sites at each scale $k$ of coarse graining.

Now, we define a renormalization transformation for the relaxation
dynamics. We will use a cell-to-site transformation on a square
lattice, in which each cell at scale $b^{(k)}$ is formed by five
sub-cells at the finer scale $b^{(k-1)}$ (see Fig. 1). We have
chosen this type of cells for two reasons: one, because it implies
the use of greater cells formed by five sub-cells at the finer
scale, that is, when we scale up, five sub-cells form a new one at
the larger scale; and second, one is intuitively tempted to follow
the geometry of the relaxation that takes place in numerical
simulations of sandpile models with energy transfer to N, E, S, W
neighbors \cite{com2}.

The length scaling factor is then
$\frac{b^{(k)}}{b^{(k-1)}}=\sqrt{5}$ (see Fig.1). Therefore, at a
generic scale $b^{(k)}$, each cell is characterized by an index
$\alpha$, ranging from one to five, indicating its number of
critical sub-cells at the smaller scale $b^{(k-1)}$. In order to
ensure the connectivity properties of the avalanche in the
renormalization procedure, only those configurations with three or
more sub-cells at scale $b^{(k-1)}$ can span the cell,
transferring energy to $i$ neighboring cells. Thus, the starting
relaxation processes $p_i^{(k-1)}$ at scale $b^{(k-1)}$ are
renormalized in the correspondent process $p_i^{(k)}$ at scale
$b^{(k)}$. Besides, it has been shown \cite{gm} that site
correlations are averaged out in the stationary state. Therefore,
taking into account this fact and the spanning rule we can write
the weight of each configuration $\alpha$ in the stationary state
as:
 \bea
 W_{(\alpha=3)}=2\rho^{3}(1-\rho)^{2} \nn \\
 W_{(\alpha=4)}=4\rho^{4}(1-\rho) \label{three} \\
 W_{(\alpha=5)}=\rho^{5} \nn
 \eea
Eq.\ (\ref{three}) gives the probability that a cell at scale
$b^{(k)}$ has the corresponding number of critical sub-cells at
scale $b^{(k-1)}$.

As an example of the general procedure, in Fig.2 we have drawn a series of relaxation
processes $p_1\rightarrow p_1\rightarrow p_2$ at scale $b^{(k-1)}$ that contributes to
the renormalization of $p_{1}^{(k)}$ at the larger scale $b^{(k)}$, starting from a
configuration of $\alpha=3$ critical sub-cells. The process consists of the following
relaxation events that span the cell from left to right satisfying the spanning
condition. First, the unstable sub-cell on the left relaxes toward the other critical
sub-cell (the center one, Fig.2b) which occurs with probability $(1/4)p_{1}^{(k-1)}$,
where the index $(k-1)$ denotes that the relaxation takes place at scale $b^{(k-1)}$.
Second, we consider the process in which the new unstable sub-cell also relaxes toward
the sub-cell on the right through another $p_1$ process (Fig.2c) which again happens with
a probability $(1/4)p_{1}^{(k-1)}$. Finally, the sub-cell on the right has become
unstable and transfers with probability $(2/3)p_{2}^{(k-1)}$ two units of energy one
inside and one outside the original cell of size $b^{(k)}$ (Fig.2d). The series of
processes described contributes to the renormalization of $p_{1}^{(k)}$. Nevertheless, it
is necessary to note that the relaxations displayed in Fig 2a-2d are not all the
processes that contribute to the renormalization of $p_{1}^{(k)}$ through a
$p_1\rightarrow p_1\rightarrow p_2$ series. Fig.2e shows a $p_2$ relaxation event that,
although involves two neighboring sites outside the original cell of size $b^{(k)}$, also
contributes to the renormalization of $p_{1}^{(k)}$ with probability
$(1/6)p_{2}^{(k-1)}$. This is a new characteristic inherent to the cell-to-site
transformation chosen. Now, if we take into account all the processes that lead to
$p_{1}^{(k)}$, for $\alpha=3$, one gets
 \bea
 p_{1}^{(k)}=\frac{1}{3}\left\{\left(\frac{1}{6}p_{2}^{(k-1)}+\frac{1}{2}p_{3}^{(k-1)}+p_{4}^{(k-1)}\right)\left(\frac{1}{4}p_{1}^{(k-1)}\right)\left(\frac{3}{2}p_{1}^{(k-1)}+\frac{4}{3}p_{2}^{(k-1)}+\frac{1}{2}p_{3}^{(k-1)}\right)\right\} \nn \\
 +\frac{2}{3}\left\{\left(\frac{1}{4}p_{1}^{(k-1)}+\frac{1}{2}p_{2}^{(k-1)}+\frac{3}{4}p_{3}^{(k-1)}+p_{4}^{(k-1)}\right)\left(\frac{1}{4}p_{1}^{(k-1)}\right)\left(\frac{3}{4}p_{1}^{(k-1)}+\frac{7}{6}p_{2}^{(k-1)}+\frac{1}{2}p_{3}^{(k-1)}\right)\right\}
 \label{four}
 \eea
where in Eq.\ (\ref{four}) the factors $\frac{1}{3}$ and $\frac{2}{3}$ refer to the
multiplicities of the configurations (see Fig.3).

In a similar way (though much more complicated), one obtains
expressions for $p_{i}^{(k)}$, $(i=2,3,4)$, for $\alpha=3$ and
imposes the normalization condition $\sum_{i=1}^{4}p_{i}^{k}=1$.
The procedure is repeated taking into account the configurations
with $\alpha=4$ and $\alpha=5$ critical sites and the renormalized
probabilities at level $k$ are finally derived by averaging over
the configurations of different $\alpha$-values including their
statistical weights $W_{\alpha}(\rho^{(k-1)})$. Therefore, the
probabilities $p_{i}^{(k)}$ at length scale $b^{(k)}$ will be
given by
 \be
 p_{i}^{(k)}=\sum_{\alpha=3}^{5}W_{\alpha}(\rho^{(k-1)})p_{i}^{(k-1)}(\alpha)
 \label{five}
 \ee
with $W_{\alpha}(\rho^{(k-1)})$ and $\rho^{(k-1)}$ given by Eq.\ (\ref{three}) and Eq.\
(\ref{two}), respectively. As the computation of the probabilities $p_{i}^{(k)}$ in Eq.\
(\ref{five}) is rather lengthy and cumbersome, we have developed a C-code to compute all
the polynomial term coefficients that contribute to the renormalization transformation.

Now, we proceed to explore the scale-invariant behavior of the model by finding the
fixed-point solution $p_{i}^{(k-1)}=p_{i}^{(k)}$. In order to do this, we start from the
shortest length scale characterized by $(\rho^{(0)},\overrightarrow{p^{(0)}})$ and study
how it evolves under repeated iteration of the transformation\ (\ref{five}). For the two
state model of Manna \cite{nsm} the parameters $(\rho^{(0)},\overrightarrow{p^{(0)}})$
are $(\rho^{(0)},0,1,0,0)$ whereas for the BTW sandpile we have $(\rho^{(0)},0,0,0,1)$.
Here, the initial value of the density of critical sites $\rho^{(0)}$ is irrelevant for
the dynamics since the system evolves to a fixed point whatever be the distribution of
critical sites at the small scale dynamics.

As in Refs \cite{pietro94,vesp95}, both models have the same fixed point, but here there
is an important difference in relation to the value of the fixed point. We obtain for the
fixed point the value $(\rho^{\ast},\overrightarrow{p}^{\ast})=(\frac{1}{4},0,0,0,1)$,
that is, in the BTW model one starts from the fixed point! This is not indeed the case
for the two-state model of Manna for which we need to iterate  Eq.\ (\ref{five}) more
than twenty times to reach the same fixed point. We believe that this is a consequence of
our renormalization strategy for the relaxation dynamics and constitutes a great
simplification in the calculation of the dynamical exponent $z$. In fact, we were
expecting the existence of a critical fixed-point value different from that reported in
references \cite{pietro94,vesp95} although the critical exponents should be very close
since they are determined by the properties of the system at large scales.

The exponent $\tau$ that characterizes the power-law avalanche
size distribution can be obtained following the procedure of
\cite{vesp95}. Consider the probability $K_{b^{(k-1)},b^{(k)}}$
that the relaxation processes that are active at scale $b^{(k-1)}$
do not extend beyond the larger scale $b^{(k)}$. This is expressed
as \cite{vesp95}
\begin{equation}
K=\frac{\int\limits_{b^{(k-1)}}^{b^{(k)}}P(r)dr}{\int\limits_{b^{(k-1)}}^{\infty}P(r)dr}=1-\left(\frac{b^{(k)}}{b^{(k-1)}}\right)^{2(1-\tau)}=1-\left(\sqrt{5}\right)^{2(1-\tau)}.
\label{six}
\end{equation}

Eq.\ (\ref{six}) also satisfies
\be
 K=p_{1}^{\ast}(1-\rho^{\ast})+p_{2}^{\ast}(1-\rho^{\ast})^{2}+p_{3}^{\ast}(1-\rho^{\ast})^{3}+p_{4}^{\ast}(1-\rho^{\ast})^{4}
 \ee
 Then, the exponent $\tau$ is given by
 \be
 \tau=1-\frac{1}{2}\frac{\log(1-K)}{\log(\sqrt{5})}=1.235 \quad.
 \ee
This value of $\tau$ is in very good agreement with the value
obtained in \cite{pietro94,vesp95} and with large-scale numerical
simulations which give  $\tau=1.27$ for the two-state model of
Manna, and $\tau=1.29$ for the BTW sandpile model \cite{luetal}.

A second independent critical exponent can also be computed. This is the so-called
dynamical exponent $z$ that relates the spatial scale $r$ to time scale $t$ through the
power law $t\sim r^{z}$. As pointed out in \cite{vesp95}, the calculation of $z$ could be
an enormous and laborious task because the knowledge of the fixed point value is not
sufficient and we have to know the complete form of the renormalized dynamics.
Nevertheless, as we said before, the use of our larger cells in the renormalization
transformation leads to a fixed point with a unique nonzero component in the vector
$\overrightarrow{p}^{\ast}$. This constitutes a great simplification in the derivation of
the complete structure of the renormalized dynamics. In what follows, we will derive at a
glance the dynamical critical exponent for the BTW sandpile model. In order to obtain the
dynamical exponent we have to calculate the average number $<t>$ of noncontemporary
processes at scale $b^{(k-1)}$ needed to have a relaxation process at the larger scale
$b^{(k)}$, which is related with $z$ through
 \be
z=\frac{\ln<t>}{\ln\left(\frac{b^{(k)}}{b^{(k-1)}}\right)}=\frac{\ln<t>}{\ln\left(\sqrt{5}\right)}
\quad .
 \label{seven}
 \ee
In Fig.3 we have depicted the possible starting configurations for the different values
of $\alpha$. It is also shown the time steps needed to have a relaxation process at the
larger scale. Such a simplification in the calculus is possible because we have to
consider only the relaxations that contribute to the renormalization of $p_{4}$ at larger
scale. As can be seen, we need two time steps for the symmetric configurations (those in
which the initial unstable site is located at the center of the cell) and three for the
non-symmetric configurations (those in which the initial unstable site is located in one
of the critical boundary sites of the cell). Therefore, \be
 <t>=\frac{1}{\sum\limits_{\alpha}W_{\alpha}(\rho)}\sum\limits_{\alpha}t^{\prime}(\alpha)W_{\alpha}(\rho)
 \label{eight}
 \ee
where $t^{\prime}(\alpha)$ is the weighted average of time steps taking into account the
different additional statistical weights due to multiplicities $\omega$ in each
configuration $\alpha$ (see Fig.3). Now, evaluating Eq.\ (\ref{eight}) at the fixed point
we obtain,
 \be
 z=1.236 \label{nine}
 \ee
The value\ (\ref{nine}) is in remarkably good agreement with the
numerical result $z=1.21$ \cite{nsm} and with the exact value
$z=5/4$ \cite{majumdar}. The other critical exponents can be
derived from scaling relations \cite{cfj}. Table I summarizes the
values of the critical exponents obtained for the BTW sandpile
model and those reported by previous renormalization scheme and
numerical simulations.

In this report, we have introduced an alternative renormalization
strategy that simplifies the analytical derivation of the critical
exponents that characterize the dynamics of sandpile models. By
using larger cells, formed by five sub-cells of the finer scale,
we obtain a fixed point with a unique nonzero dynamical component
which allow us to derive the whole form of the renormalized
dynamics in a more direct and simple way. The values of the
exponents here obtained are in good agreement with those
previously reported. Besides, as in similar analytical
predictions, the two-state model of Manna and the BTW sandpile
model belong to the same universality class \cite{com3}. The
results confirm the robustness of the renormalization group
approach.

It is a pleasure to thank A. Vespignani for stimulating
discussions and G. Caldarelli for useful correspondence. Y.M would
like to thank the AECI for financial support.

%\end{multicolumns}
%**************************************
%**************************************

\begin{table} \caption{Values of the critical exponents for the BTW sanpile model
$(d=2)$. We have included the values obtained from large scale
simulations and those derived in [6].}
\begin{tabular}{ddddd} Method&$\tau$&$\alpha$&$\lambda$&$z$\\
\tableline RG\cite{vesp95}&1.253&1.432&1.506&1.168\\
Simulations\cite{luetal,nsm}&1.29&1.38&1.44&1.21\\ This
paper&1.235&1.38&1.47&1.236\\
\end{tabular} \label{table1} \end{table}

\newpage

\begin{figure}
 \caption{Greek cross-shaped cell used in the renormalization procedure. It
is displayed the central sub-cell (encircled dot) and its four
nearest neighbors (black dots). The length scaling factor is
$\sqrt{5}$.}
%\epsfxsize=8.6cm
%\epsfysize=8.6cm
%\epsfbox{fig5.eps}
\label{figure1}
\end{figure}

\begin{figure}
 \caption{A series of relaxation processes $p_1\rightarrow p_1\rightarrow
p_2$. Open dots represent stable sites, filled dots critical
sites, and encircled dots unstable sites. Note that the last
relaxation affects only one neighbor despite of having two outward
arrows (see also Fig.1).}
%\epsfxsize=8.6cm
%\epsfysize=8.6cm
%\epsfbox{fig5.eps}
\label{figure2}
 \end{figure}

\begin{figure}
 \caption{Full set of possible initial configurations of critical sites and
their multiplicities $\omega$. We have only depicted the
configurations that fulfill the spanning rule $(\alpha=3,4,5)$.
The t's refer to the noncontemporary time steps needed to have a
relaxation that span the whole cell. The indices $s$ and $a$ stand
for symmetric and non-symmetric configurations.}
%\epsfxsize=8.6cm
%\epsfysize=8.6cm
%\epsfbox{fig5.eps}
\label{figure3}
 \end{figure}


\begin{references}
\bibitem{note1} These cases have been recently considered and do
not lead to significant differences in the values of critical
exponents.
\bibitem{com2} Strictly speaking, the choice of the Greek
cross as basic renormalization cell, does not ensure that the
answers are better.
\bibitem{com3} It is worth at this point to recall that there are
several recent preprints that report that the model of Manna is in
a different universality class than the BTW model. These preprints
are based on extensive numerical calculations.
\end{references}
\end{document}